# UNIVERSALIZING SCIENCE: ALTERNATIVE INDICES TO DIRECT RESEARCH

Ari Melo Mariano* e Maíra Rocha Santos**

## ABSTRACT

Measurement is a complicated but very necessary task. Many indices have been created in an effort to define the quality of knowledge produced but they have attracted strong criticism, having become synonymous with individualism, competition and mere productivity and, furthermore, they fail to head science towards addressing local demands or towards producing international knowledge by means of collaboration. Institutions, countries, publishers, governments and authors have a latent need to create quality and productivity indices because they can serve as filters that influence far-reaching decision making and even decisions on the professional promotion of university teachers. Even so, in the present-day context, the very creators of those indices admit that they were not designed for that purpose, given that different research areas, the age of the researcher, the country and the language spoken all have an influence on the index calculations. Accordingly, this research sets out three indices designed to head science towards its universal objective by valuing collaboration and the dissemination of knowledge. It is hoped that the proposed indices may provoke new discussions and the proposal of new, more assertive indicators for the analysis of scientific research quality.

## UNIVERSALIZACIÓN DE LA CIENCIA: ÍNDICES ALTERNATIVOS DE DIRECCIÓN DE LA INVESTIGACIÓN

### RESUMEN

La medición de la calidad es una tarea compleja. En 1972 Eugene Garfield ofrecería a la ciencia el Factor de Impacto (FI), uno de los índices para calificar el conocimiento producido más utilizado y al mismo tiempo criticado de la ciencia contemporánea. Desde entonces, otros índices para calificar la ciencia y los autores vienen apareciendo, como el h-index, CiteScore o factor de imediatismo. los índices actuales se preocupan en evaluar la calidad y la productividad científica, pero no conducen la ciencia para atender las demandas locales. En esta perspectiva este estudio busca responder: ¿es posible ofrecer nuevos indicadores para la ciencia, aprovechando la contribución de los índices existentes y dirigiendo la investigación para un desarrollo de la comunidad científica en su conjunto y no a favor de los intereses individuales? Así el objetivo de esta investigación es presentar tres índices de direccionamiento de la ciencia para un objetivo universal, valorando la colaboración y la dispersión del conocimiento. Se espera, con los índices propuestos levantar nuevas discusiones y proposiciones de indicadores más asertivos sobre la investigación científica.

**Palabras Clave:** Factor de impacto, índices bibliométricos, índices alternativos, bibliometría, ciencia.

## FUENTES DE FINANCIACIÓN / FUNDING
Coordenação de Aperfeiçoamento de Pessoal de Nível - CAPES

* arimariano@unb.br - Universidade de Brasília – Brasil / University of Brasília – Brazil

** mairarocha@unb.br - Universidade de Brasília – Brasil / University of Brasília – Brazil

**INTRODUCTION**

Measuring quality is a complicated task. In 1972, Eugene Garfield offered the scientific world his Impact Factor, one of the most-used indices in the world for qualifying knowledge produced by contemporary science and for criticizing it too (Moed, 2010). Garfield (1973) himself made it clear that his indicator could not be used in the same way for different research fields because of various factors that influence its calculation. Pinski and Narin (1976), Moed et al. (1996a, 1996b, 1998) pointed out that the age or aging of the journal could easily introduce bias when using it to qualify science just as Moed and Van Leeuwen et al. (1996) declared that the time taken to locate a document citation varies from journal to journal, thereby jeopardizing the final index obtained. Since then other indices for qualifying science have appeared such as the H-Index, and the CiteScore. The existence of these indices and their adoption by Universities and Development Agencies have transformed research into a high-productivity race and a quest to achieve the highest possible number of citations and publications in renowned reviews or journals, that is those that are recognized as having the highest Impact Factors.

However, Da Silva and Memon (2017) have made it clear that the discussion of the IF goes well beyond its inability to measure the quality of scientific production and that its interests are far more associated to those of the big publishing houses than to those of science as such. In that sense it would actually function as a symbol of the hegemony among authors, universities and countries. The vast majority of journals indexed using this system require that articles should be submitted in English and with their citations based mainly on other reviews with IF indices in the first quartile and important author h-indices. That is evident in the instructions to authors that the journals make available in their submission pages. In their critical observations on that context, Harzing and Alakangas (2016) insist on the need for a truly multi-lingual database because in their study of the Google Scholar they found that authors that appear as the most cited on a given theme in the languages of their own countries such as Brazilian Portuguese, German and French, are entirely invisible in the data of Web of Science and Scopus.

The difficulty of expressing oneself correctly when writing in a non-native language, exacerbated by the requirement of using formal technical language and the different rules applying to each publication, have led non-English language countries to create their own indices to qualify periodicals. Kulczycki and Rozkosz (2017) describe how Poland has created its own national index to rate journals that do not have an IF rating that would enable them to be present in international databases; an attempt to value their national science by using experts to assess the knowledge produced in the periodicals. In Brazil the Webqualis

(https://sucupira.capes.gov.br/sucupira/public/consultas/coleta/veiculoPublica caoQualis/listaConsultaGeralPeriodicos.jsf) does the same, indexing national and international reviews based on its own criteria. Authors like Ramirez and Mariano (2014), however, reveal that those classifications have aspects that need to be improved insofar as they attribute the same weightings to situate journals with different levels without providing an apparent explanation. Da Silva and Memon (2017) warn that one of the reasons why the Impact Factor has continued to be an international index for almost 60 years is that many of the parallel indices that have appeared were found to be making fraudulent manipulations.

Thus it can be seen that while current indices are concerned to evaluate scientific quality and productivity, they fail to direct science towards addressing local demands or the internationalization of knowledge; they have become measures centered on the individual and/or the periodical, but not on science itself. In that perspective, the present study aims to address the question: is it possible to offer new science quality indices, making use of the contribution of existing ones and directing research towards the development of the scientific community as a whole and not merely to favoring individual interests?

Thus this research sets out to present three indices intended to head science towards a universal objective that values collaboration and the diffusion of knowledge.

## LITERATURE REVIEW

### Impact Factor

Glänzel and Moed (2002) state that despite the criticism it receives, the Impact Factor is still the most used index for measuring scientific quality and it is calculated in two-year intervals using Equation (1),

$$IF_n = \frac{C_{(n-1)+(n-2)}}{P_{(n-1)+(n-2)}} \ (1)$$

Where IF is the impact factor, n is the year, C is the number of citations and P is the number of publications. Thus the Impact Factor for 2017 is equal to the number of citations of the periodical's articles in 2016 (2017 − 1) plus the number of citations of the periodical's articles in 2015 (2017 − 2) divided by the quantity of publications in the same periodical in 2016 (2017 – 1) plus the number of articles published in the periodical in 2015 (2017 – 2).

Kulczycki and Rozkosz (2017) explain that decision-making on research financing, career promotions and salary scales generally requires and makes use of IF ratings. However, counterbalancing that practice, in December 2012 in San Francisco, USA, the Declaration on Research Assessment (DORA) was published making important recommendations for research assessment. Among them were: not basing evaluation of research investigations on Impact Factor alone; considering the content and not merely the metrics when evaluating researcher's work; and adopting other practices to qualify researchers, especially junior researchers. Similarly, in 2016 in the Bolivian city of Sucre, Bolivia, Latin American researchers created the Southern Axis Social Scientists Network (*Científicos Sociales del Eje Sur – RedeCSES*) calling for research evaluation to be based on its social impact on local communities and to value that kind of scientific knowledge over research in which secondary data are the main inputs and researchers conduct their activities shut up in their offices, often without ever coming face to face with their study objects. Both those manifests clearly state that they have no intention of dislodging the Impact Factor but rather of tempering it with other variables capable of favoring different research perspectives

In spite of the challenge the IF presents, many other metrics have appeared in an effort to cast more light on the question of qualifying research, authors and periodicals and one of the best-known is the h-index.

**H-index and CiteScore**

Hirsch proposed the h-index in 2005 and it is a cross between quality represented by the number of citations and the quantity of publications. Authors like Vanclay (2008), Martín, (1996) and Alonso, et al. (2009) are enthusiasts of the index; they consider its main advantage to be its easy calculation and the fact that the h-index can reflect an author's stability.

Over the years, many authors (Egghe, 2006, Bertoli-Barsotti and Lando, 2017; Bornmann, et al. 2008 & Jin, 2007) have sought to adapt the h-index while others, like Carter; Smith, and Osteen (2017), have used to it make comparisons regarding gender and position held in the university. Figure 1 illustrates how the h-index is calculated.

**Figure 1.** Number of citations by the number of published articles.

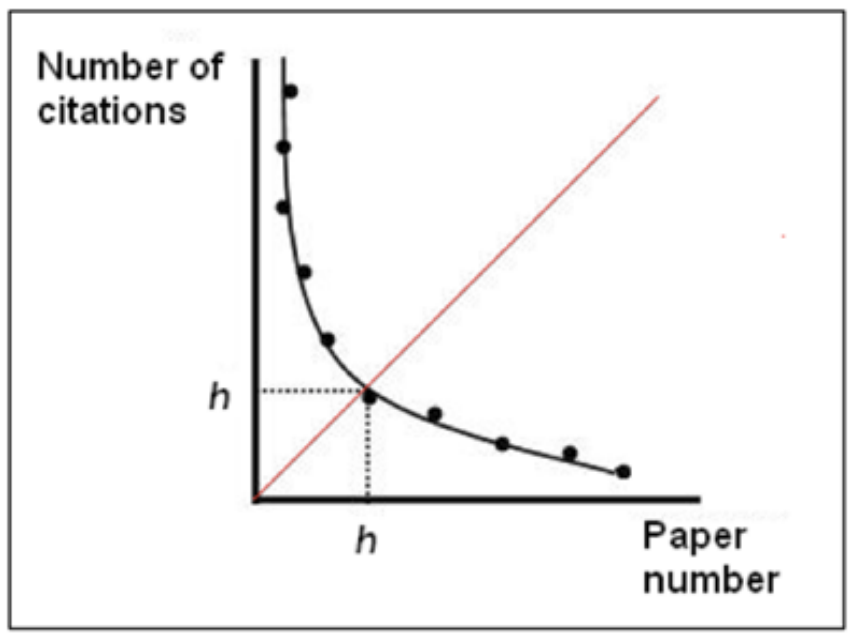

Source: Yin, (2012), adapted from Hirsch (2005).

The figure shows how the intersection of the 45º line with the curved line indicates the h-index. Thus for an author to obtain an h-index of 4, he would need to have had at least 4 articles published and each article would need to have been cited at least 4 times; an author with an h-index of 20 would need to have had 20 articles published with each one having been cited 20 times. This metric diminishes the situation of an author with 50 articles published and only one cited, but cited 1,000 times because his average citation would be 20 citations per article but his h-index would be just h1. Currently the Google Scholar automatically calculates the author h-index of those researchers who register with their online platform.

Although the h-index has been less criticized than the IF, it too has limitations such as only being applicable to authors in a same field or the difficulties associated to tallying the citations of authors with very similar names (Hirsch, 2005).

A more recent effort to obtain a better, fairer index for measuring research quality was officially created in December 2016, namely the CiteScore (CS) index.

Da Silva and Memon (2017) consider that the CS presents some important differences from the IF; the CS belongs to a publishing house, uses a three-year interval, includes articles, reviews, editorials and articles presented in conferences, it is free of charge (/journalmetrics.scopus.com) and is updated on a monthly basis. The IF does not belong to a publisher, uses a 2-year interval and only accepts journal articles; it requires subscription and its updates are annual. What is concerning is the insistence that while the IF is influenced by factors associated to editorial policy, the CS is not and that supposedly endows it with transparency. The calculation of the CS is very similar to that of the IF as shown in Equation (2):

$$CS_n = \frac{C_{(n-1)+(n-2)+(n-3)}}{P_{(n-1)+(n-2)+(n-3)}} \ (2)$$

Where CS is the Citescore, C is the number of citations and P is the number of publications. Thus the Citescore for 2017 is equal to the number of citations of the periodicals articles in 2016 (2017 − 1) plus the number of citations of the periodical's articles in 2015 (2017 − 2) and 2014 (2017 − 3), divided by the quantity of publications in that same periodical in 2016 (2017 − 1) plus the number of articles published by the periodical in 2015 (2017 − 2) and 2014 (2017 − 3)

Van Noorden (2016) insists that the criteria adopted by the CS are fairer but what is causing concern and uneasiness in editorial circles, especially those of the big publishing names like Nature and Science, is that it is still too early to properly evaluate this new metric, especially because the IF, the h-index and the CS are all equally orientated towards measuring the quality of research, periodicals and authors based on productivity criteria to the detriment of a prerogative that is highly important to science: direction.

The direction referred to here is towards the overriding goal of science and the main reason for its existence: the development of society and its environment. A consultation of the Scimago Journal & Country Rank (SJR), 2017 database with information on the progress of science among countries showed that in the preceding 30 years (1996-2016), there were no significant changes in the research quality ratings and positions of the United States, the United Kingdom or Germany, according to indices like the h-index or citations per document as shown in Figure (2). It is also clear that that while in 1996 China was not among the top four countries in research ratings (it was in ninth position), by 2006 it was in second place, surpassed only by the United States. An analysis of the self-citations as a percentage of the total citations revealed that China made use of its continental population size to gain volume insofar as 54.11% of all citations were self-citations. However, a comparison of the h-index with those of the other countries shows that China's documents failed to achieve a balance between quantity (quantity of publications) and quality (quantity of citations).

**Figure 2.** H-index, % of self-citation and citation per document graphs 1996, 2006, 2016.

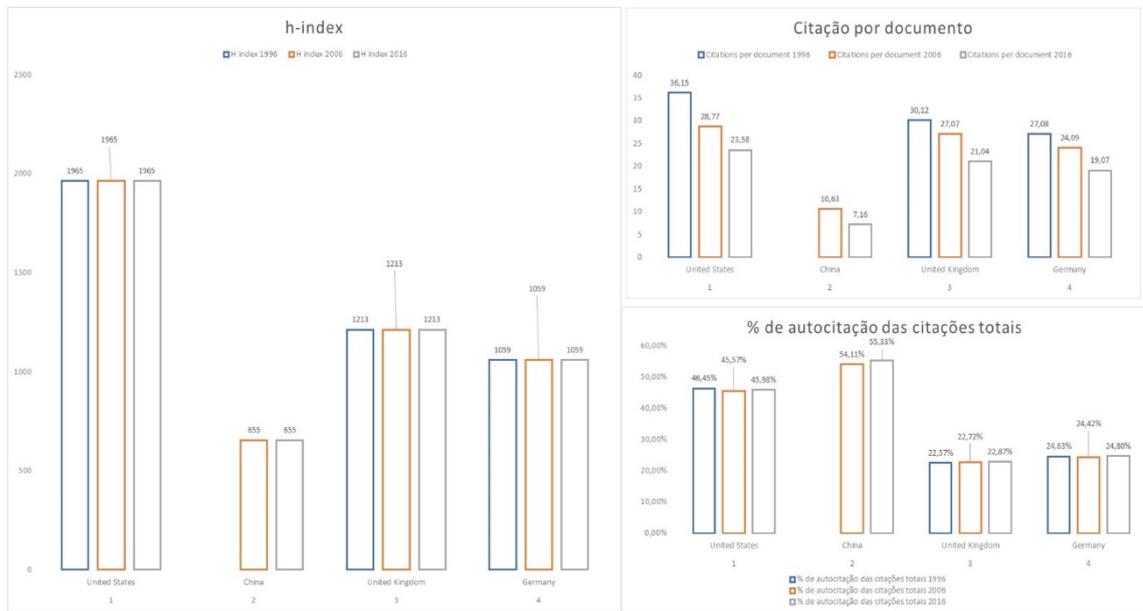

Source: SJR (2017).

Observing scientific production beyond the bounds of that select group (United States, China, United Kingdom, Germany) reveals that countries that do not appear in the graphs, like Brazil, show very little movement according to the analyses carried out and always occupy positions from the 13[th] to the 15[th] place. That observation is important insofar as it brings up the question of direction because, to date, almost all the indices presented as a means of qualifying research are actually instruments to reaffirm the productivity of a hermetic group of already consecrated authors and publishers. Thus there is a need to offer other feasible, transparent options capable of making good use of the formerly created structures to foster a science that is at once fairer and more universal.

**METODOLOGY AND RESULTS: ALTERNATIVE INDICES FOR QUALIFYING AND DIRECTING SCIENCE**

Given the present day context regarding quality and productivity indices for science, this study has opted to put forward some suggestions designed to foster a gnosis directed at the development of a universal research ecosystem.

**Hcol -Index(collaboration h-index)**

The hcol is an attempt to foster collaboration among researchers in different levels and favor the growth of the scientific community as a whole. The first step must be to establish the basis for collaboration. The rules are established by taking the h-indices of two researchers, Author 1 and Author 2. Author 1 is the more experienced researcher and so he has a higher h-index while author 2, being less experienced, has a lower one. As an example we can suppose that collaborating author 1 has an h-index of 40. To discover with whom he should collaborate, that author must multiply his own h-index by 0.1 and so he is destined to collaborate with a second author who has an h-index of 4 as illustrated in Figure (3).

**Figure 3.** Calculation for collaboration between h-indexed authors

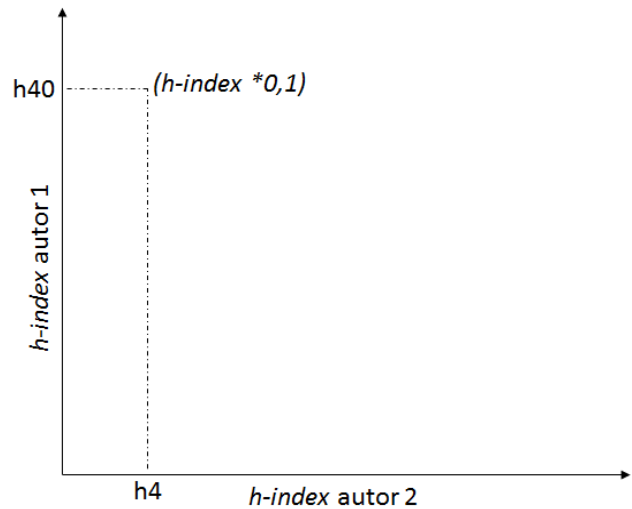

Source: author of this paper

Once the rules governing choice of collaborators have been established the $h_{col}$.index can be calculated using Equation (3).

$$h_{col=} \frac{Qt_{tpcol2}}{Qt_{ttpcol1}} \ (3)$$

Where hcol is the h-index for collaboration and is given by dividing the quantity of studies undertaken jointly by collaborator 2 and collaborator 1 ($Qt_{tpcol2}$), by the total number of studies undertaken by collaborator 1 alone ($Qt_{ttpcol1}$).).

The results are advantageous for both authors (1 and 2) because the more experienced author increases his score in the collaboration index and has access to different perspectives and new academic frontiers, while the less experienced author  improves his penetration in the academic world  and learns how to formulate a high-quality paper. The gains for science can be considerable because collaboration between authors situated in different academic realities fosters human relationships among researchers as well as interdisciplinarity. In practice this collaboration index is an application of the graphs shown earlier to the aspect of scientific collaboration but not from the passive perspective of understanding who is collaborating but instead, in the light of who should be collaborating with one another. Researchers normally operate in a highly modular network of dense connections among their peers but very sparse connections with researchers in other groups or modules. Thus the aspect of modularity offers the means to identify possible, promising relations (Newman, 2006). In that perspective science will almost always be limited by the group to which the researcher belongs; authors with an h-index in the range of 40 to 50 will tend to relate to others in the same category and that limits their investigations to a unidirectional perspective. The moment a researcher with an h-index of 4 or 5 starts to collaborate with a group in the h-40 to h-50 range a new dimension opens up, not only for the further development of science but also for social and cultural development. Figure (4) illustrates that kind of relation.

**Figure 4.** Examples of $h_{col}$ advantages

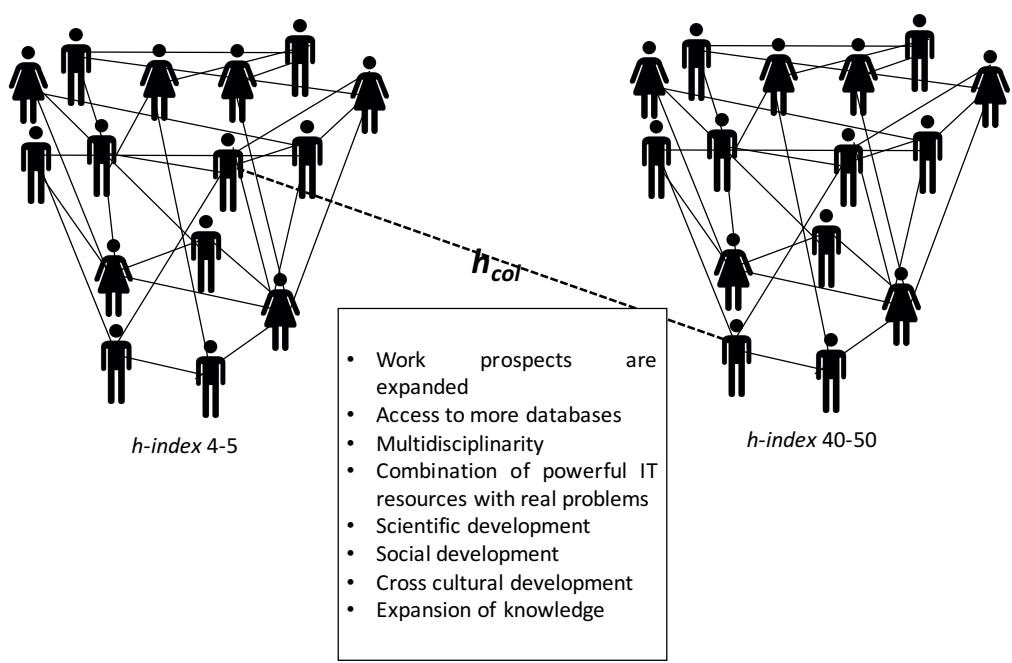

$h_{col}$

h-index 4-5

h-index 40-50

- Work prospects are expanded
- Access to more databases
- Multidisciplinarity
- Combination of powerful IT resources with real problems
- Scientific development
- Social development
- Cross cultural development
- Expansion of knowledge

Source: author of this paper

In this situation both researchers gain access to other realities, other databases and sources of information which they would otherwise find very difficult or even impossible to achieve. This kind of collaboration has been observed recently in the case of the outbreak of Zika virus infections which set the entire scientific community in collaborative action to try and solve the problem. The Brazilian government published official call Nº 14/2016 for research into combating and preventing Zika Virus. Similarly, the United States adopted a strategy and launched a call for research by means of its USAID (USAID.gov) and another by its Grants.gov platform. Even the European Union showed its willingness to foster solutions through its European Commission, launching official calls for research and providing incentives. Taking the various different health agencies that proposed alliances to find solutions for the epidemic as an example, we can conclude that the $h_{col}$ ratings could contribute with the results of collaboration, becoming a kind of 'Médecins sans frontières' of the scientific community.

**Academic Impact Index**

The Academic Impact index (AII) is an endeavor to stimulate closer relations between research professors and undergraduate students. In the current situation the figure of the university teacher is dissociated from that of the researcher as if the two were antagonistic pathways in the career of university teaching professionals (Cookson, 1987; Doig, 1994; Gurney, 1989). On the other hand, the need for a more analytical professional capable of handling great volumes of data intensifies the demand for research at the same time as it molds the profile of the individual. Zeiochner (1998) believes that only the involvement of university teachers with their undergraduates will make that kind of education feasible. Thus the AII is an index that takes up, once more the question of teacher-student relations, introducing research into the normal routine of the education tripod; research, teaching and extension. This Index is calculated by means of Equation (4):

$$IAC = \frac{qt_{ace}}{\left(\frac{qT_{as}}{id}\right)} \quad (4)$$

where the Academic Impact Index is equal to the total quantity of articles elaborated in partnership with undergraduate students (qtace), divided by total quantity of articles elaborated without the collaboration of undergraduates (qtas) divided by the age of the researcher (id). The introduction of the researcher's age is an attempt at weighting that variable. Moed et al. (1996a, 1996b, 1998) consider that it is a factor that must be taken into account. Thus a 53 year-old university teacher who has published 90 articles of which five were produced with the collaboration of undergraduates, would have an AII of 2.94 according to equation (5). Similarly Equation (6) exemplifies the case of a 31 year-old university teacher with 40 publications, of which 15 articles were produced with the collaboration of undergraduate students. It is clear that in spite of being younger and having a lower total number of publications than the researcher in the first example, his AII is 11.62 which underscores the contributive nature of the teacher-student relationship.

$$IAC = \frac{5}{\left(\frac{90}{53}\right)} \xrightarrow{r} \mathbf{2,94} \quad (5)$$

$$IAC = \frac{15}{\left(\frac{40}{31}\right)} \xrightarrow{r} \mathbf{11,62} \quad (6)$$

Thus the Index seeks to expand collaboration between university teachers and professors and undergraduates in the sense of orientating the study discipline contents towards scientific discoveries, making them more aligned with the new discoveries in each scientific field. It also creates an opportunity to improve teacher-student relations in the context of contemporary education. It must be underscored that joint publication should be done during the student's period of graduation, favoring the score of the same even when the student concludes the course, because the act of publication is the temporal marker of the collaboration.

**Responsible Social Impact (RSI) Index**

The responsible social impact index intends to draw science closer to its study objects by ensuring the researcher's commitment to the research problem. Bueno (2010) explains why there is such a great distance separating scientific communication from its dissemination. Scientific communication is directed at other scientists, that is, at the peer audience whereas dissemination endeavors to diffuse science to the lay public (Bueno, 2009). The role of scientific diffusion in the media has been carefully observed because it has the power of influencing society and, accordingly, many media entities contract scientific consultants to ensure a greater degree of assertiveness in the information they broadcast or publish or display in films (Szu; Osborne, & Patterson, 2016).

However, a researcher's commitment and obligation to science does not affirm itself merely through the diffusion of knowledge to his or her peers and to society at large; it also takes place through providing some form of direct return from the findings to those who provided the primary data or who made available channels or persons necessary to the investigation. While it is true that a review of the literature in Web of Science, Scopus and Google Academic failed to identify any results related to such research feedback, the very reading of scientific papers and professional work as a teacher and researcher ratify the fact that studies of documents that offer feedback on findings associated to a given problem are very scarce. Rather than being a mere expression of gratitude for the time dedicated to participating in the research, such gestures are an extremely valuable contribution to those professional individuals that work in the reality targeted by the study. Many institutions are closing their doors to research projects because the return they have received from collaborating with them has been so small. Accordingly this particular index seeks to mitigate the situation and imbue the research with responsibility and commitment to the study object. It is calculated by means of equation (7):

$$ISR = \left(\frac{qt_{ar}}{qt_{at}}\right) + (ad*0,1) \ (7)$$

where the Responsible Social Impact index is equal to the quantity of articles published that have registered feedback ($q_{tar}$), divided by the total quantity of articles (qtat), plus the number of actions of diffusion to society at large (ad) multiplied by (0.1).

Feedback in this situation is considered to be a letter/document/report delivered to the place where the research was undertaken in two copies one of which sets out the main results of the research signed by the person responsible for making the study feasible. Furthermore, that letter/document/report should be annexed to the published study. In this context, actions of dissemination to society at large are considered to be any dissemination initiative outside the ambit of the university directed at non-researchers. Such actions should be described with information on the place, date and time they took place and signed by the person responsible for the establishment involved or for the event itself.

This index may at first sight appear to be difficult to operationalize but it actually makes a correction to ensure fairness, that is, it stimulates the provision of feedback on the study to the object of the study and further rewards any extra actions in that sense that the researcher may eventually decide to undertake. It must be underscored that the absence of feedback is serious, insofar as it indicates the lack of commitment on the part of the researcher. Thus, in a comparison of two researchers in which researcher (a) provided return on all 10 of his research efforts but failed to promote any action for society and researcher (b) only provided return on 5 of the 10 projects he executed but nevertheless carried out 3 actions directed at society at large, then the calculations would be as shown by equations (8) and (9):

$$ISR_A = \left(\frac{10}{10}\right) + (0*0,1) \rightarrow 1,00 \ (8)$$

$$ISR_B = \left(\frac{5}{10}\right) + (3*0,1) \rightarrow 0.80 \ (9)$$

Thus it is easy to see that the proposed index is far more than a mere qualifier, it is a means of correcting a fault that is present in many papers and it orientates science in the direction of a greater nearness to society as a whole. Among its main advantages are improved relations between researchers and society, maintaining the Universities' institutional image and thereby enabling them to gain access to various different realities and to be sure that they can make a real contribution, and drawing research and extension closer together, considering that diffusion to society at large can readily take place through extension activities.

It is important to point out that the RSI index also has its restrictions insofar as not all branches of science contemplate direct feedback to society although it is possible that actions in that direction may exist in the future.

**CONCLUSIONS**

This study presented three indices designed to head science towards the universal finality of valuing collaboration and dispersal of knowledge. Namely the Collaboration h-index, the Academic Impact index and the Responsible Social Impact index. The primordial factor inherent to each one is that of heading science in the direction of universalization.

It is easy to understand institutions, countries, publishers and authors' need to create academic quality and productivity indicators insofar as they become filters for a wide scope of decisions and even decisions on the promotion of university teachers in their professional careers. However in the present-day context, the very creators of the available indices make it clear that they were not exactly designed for that purpose because in different areas of research, things like the age of the researcher, the country, the spoken language and others all influence the index calculations. That said, it is, nevertheless, necessary to make best use of the structure and hegemony of the established indices and, by means of weightings, guide science towards the creation of an ecosystem of collaboration and development.

It must be stressed that the indices presented in this research are alternative indices and they are not intended to replace those already being used. Accordingly, those parameters that have already been consolidated in the academic world have served as the references for this study. It is hoped that the indices proposed here will serve to provoke further discussion and new propositions of more assertive scientific research indicators.